%% file: FDOA_FF_main.tex
\documentclass[12pt]{amsart}
\usepackage{graphicx}
\usepackage{url,amstext,amsfonts,amssymb,amscd,amsbsy,amsmath,amsthm,verbatim,mathrsfs}
\usepackage{ifthen, fullpage, mathrsfs, cite,hyperref,framed}
\usepackage{color}
\usepackage{algorithm}% http://ctan.org/pkg/algorithm
\usepackage{algpseudocode}% http://ctan.org/pkg/algorithmicx
\usepackage{longtable}
\usepackage{rotating}

\usepackage{caption}
\DeclareCaptionFormat{myformat}{#3}
\theoremstyle{definition}

\theoremstyle{remark}

\newcommand{\mbf}[1]{\mathbf{#1}}

\begin{document}

\title{A Novel Method for Determining DOA from Far-field TDOA or FDOA}
\author{Karleigh J. Cameron }
\author{Samuel J. Pine } % I couldn't get emails/ affiliations to work- we can worry about it later

\input{abstract}

\maketitle

%%%%%%%%%%%%%%%%%%%%%
%%%%%%%%%%%%%%%%%%%%%
%%%%%%%%%%%%%%%%%%%%%
% Intro / background / lit review ?
\input{background}

%%%%%%%%%%%%%%%%%%%%%
% FDOA derivation
\input{FDOA}

%%%%%%%%%%%%%%%%%%%%%
% TDOA derivation
\input{TDOA}

%%%%%%%%%%%%%%%%%%%%%
% Algorithm description
%\input{algorithm}

%%%%%%%%%%%%%%%%%%%%%
% Numerical results
\input{numericalresults}

%%%%%%%%%%%%%%%%%%%%%
% Future work & conclusion
\input{conclusion}

%\subsection{Acknowledgments}

\bibliographystyle{abbrv}
\bibliography{fdoa.bib}
\end{document}

%% file: abstract.tex
\begin{abstract}
Passive source localization is often performed using time difference of arrival (TDOA) measurements, frequency difference of arrival (FDOA) measurements, direction of arrival (DOA) measurements, or a combination of all of these. For a source in the far-field, DOA can be extracted from the TDOA and FDOA measurements due to simplifications that arise in the far-field approximation. This paper presents this relationship and the corresponding DOA estimation method. Utilizing TDOA and FDOA measurements for computation of signal DOA requires only a linear solve, which makes the corresponding source localization technique very efficient. Additionally, the method provides an inherent de-noising of receiver measurements, since they are being projected onto the range of the receiver differencing matrix.
\end{abstract}

%% file: background.tex
\section{Introduction}
\label{s:intro}
Locating a radio-frequency transmitter, or \emph{source localization}, is a vital step in many applications. Source localization is often performed using measurements of the transmitted signal obtained by several nearby receivers. Specifically, measurements of the transmitted signal at two distinct receivers allow one to compute the time difference of arrival (TDOA) and frequency difference of arrival (FDOA) between those receivers. With estimates of TDOA or FDOA measurements, one can compute various other quantities describing the location of the transmitter, including the angle of arrival (AOA) / direction of arrival (DOA), the range to the receiver, and thus the location of the transmitter in the global coordinate system (geolocation). If information about the source is known {\em a priori}, such as altitude (ALT), it is typically possible to estimate receiver location with the use of fewer measurements.\\

Source localization using only TDOA measurements is a well-understood problem and many algorithms have been developed for its solution. Common approaches include linearization of the system or a multidimensional search~\cite{Torrieri1984}. Methods for managing data to deal with noise, including divide and conquer (DAC), the RANdom SAmpling Consensus method (RANSAC), and projection to the feasible set of TDOA measurements have been proposed ~\cite{Cameron,Abel1990,Li2009,Compagnoni2017}. Geometrically, each TDOA measurement restricts the potential transmitter location to a hyperboloid. Thus, if several measurements are obtained, locating the emitter requires finding the intersection of several hyperboloids. Simple geometric relationships between the TDOA measurements and the known receiver positions allow the DOA to be computed with a single antenna array~\cite{Benesty2008}. It follows that with multiple antenna arrays, the source can also be located via triangulation. \\

The equations relating FDOA measurements to receiver positions are more complicated than the corresponding TDOA equations. The FDOA model is nonlinear and depends on the receiver velocities, so source localization with FDOA measurements is more complicated than geolocation using TDOA measurements. Additionally, since FDOA measurements quantify the Doppler effect between receivers, it is essential that each receiver has a different velocity. This makes FDOA approximation with an antenna array impossible. \\

While the FDOA measurements are often used as an additional constraint to the TDOA geolocation systems (TDOA/FDOA localization)~\cite{Ho1997}, only a few algorithms have been developed using FDOA alone~\cite{Cameron,Jinzhou2012}. Some limitations of these algorithms are the high cost of computation that comes from nonlinear solver methods. There are, however, some cases where it is desirable to solve for the emitter location using FDOA only. For instance, in the case of a narrowband signal with a long pulse duration, Doppler resolution is finer than the range resolution so that it is difficult to measure the TDOA accurately~\cite{Cheney2009,Mason2005,Jinzhou2012}. \\

When the distance between the receivers and the transmitter is much greater than the distance between the receivers it is common to simplify the wave propagation model and assume that wave curvature is negligible in the region of the receivers. This assumption is commonly referred to as the far-field assumption~\cite{Cheney2009}. In this paper we present how this assumption can reduce the computation of DOA to the solution of a linear system. \\

While DOA estimation is typically performed with a TDOA-based strategy, our approach is able to utilize TDOA or FDOA measurements, or both, by capitalizing on the simplified geometry of the source-localization problem under the far-field assumption. One scenario where this method might be useful is in the calculation of DOA of a narrowband emitter using several receivers. The main benefit of this method is its computational efficiency, as it simplifies the calculation of DOA to solving a linear system of equations.  With several DOA calculations, triangulation can be used to determine location of the source. In section~\ref{s:FDOA}, we develop a far-field model for the FDOA measurements and discuss a technique for determining the signal direction of arrival. In section~\ref{s:TDOA}, we develop a similar far-field approximation for the TDOA model and present the analogous DOA technique. Finally, we summarize the method with some numerical results in section~\ref{s:numerics} and concluding remarks in section~\ref{s:conclusion}.

%% file: FDOA.tex
\section{Direction of Arrival with FDOA Measurements}
\label{s:FDOA}

Consider a stationary transmitter located at $\mbf{x}$. Suppose we have $N$ receivers located at $\mbf{x}_1, ...,\mbf{x}_N$ with velocities $\mbf{v}_1, ...,\mbf{v}_N$. The frequency shift of the signal between the emitter and the $i^{th}$ receiver is
\begin{align}
  \label{eq:fshift}
  d_i =  \frac{f_0}{c}\left(\mbf{v}_i^T\cdot\frac{\mbf{x}_i-\mbf{x}}{\|\mbf{x}_i-\mbf{x}\|}\right),
\end{align}
where $f_0$ is the emitted frequency and $c$ is the speed of wave propagation in the media. In our scenario, we know the receiver positions and velocities, and we would like to solve for the transmitter position. We cannot measure the frequency shift directly, but we can measure the difference in frequency shifts,
$f_{i,j} = d_j-d_i$. Scaling by $\|\mbf{x}_i-\mbf{x}\|^{-1}$ means that each of these equations is nonlinear. Even so, by taking pairwise differences of the equations in \eqref{eq:fshift}, one can numerically solve the system by expressing it in terms of polynomials and using techniques like homotopy continuation \cite{Cameron}. While such an approach is able to find all solutions, it is computationally more expensive than a linear solve.

The nonlinearity of \eqref{eq:fshift} makes it difficult to accurately solve for $\mbf{x}$, so we simplify the model by deriving a far-field approximation for Equation \ref{eq:fshift}. Additionally, for simplicity we ignore the constant factor of $f_0/c$ in~\eqref{eq:fshift}. Thus $d_i$ is now \textit{proportional} to the frequency shift.

\subsection{Far-field Approximation for FDOA}
Assume without loss of generality that the receivers are centered around the origin. We consider the far-field case, where the distance between receivers is much smaller than the distance to the emitter, i.e. $\|\mathbf{x}-\mathbf{x}_i\|>>\|\mathbf{x}_i\|, \; \forall \; i$. The far-field approximation (as in~\cite{Cheney2009}) for $1/\|\mathbf{x-x_i}\|$ is:
\begin{align*}
  \frac{1}{\|\mathbf{x-x_i}\|} = \frac{1}{\|\mathbf{x}\|}\left(1+\mathcal{O}\left(\frac{\|\mathbf{x}_i\|}{\|\mathbf{x}\|}\right)\right).
\end{align*}
Truncating after the first term above allows for simplification of the factor (in eq. \ref{eq:fshift}):
\begin{align*}
  \frac{\mbf{x}_i-\mbf{x}}{\|\mbf{x}_i-\mbf{x}\|} \approx \frac{\mbf{x}_i}{\|\mbf{x}\|}-\frac{\mbf{x}}{\|\mbf{x}\|}.
\end{align*}
Additionally, the far-field assumption implies that the first term will have small magnitude. Thus, $\dfrac{\mathbf{x_i-x}}{\|\mathbf{x_i-x}\|}$ is simplified to $\dfrac{-\mathbf{x}}{\|\mathbf{x}\|}$.
Equation \ref{eq:fshift} becomes:
\begin{align}
  d_i =  -\mbf{v}_i^T\cdot\hat{\mbf{x}},
\end{align}
where $\hat{\mbf{x}} = \frac{\mbf{x}}{\|\mbf{x}\|}$, is the unit vector in the direction of $\mbf{x}$. The entire system of frequency shifts can be written:
\begin{align}
  \label{eq:fshiftFF}
\mbf{d} = -\mbf{V}\hat{\mbf{x}},
\end{align}
where \begin{align*}
\mathbf{d}=\begin{pmatrix}
d_1 \\ \vdots \\ d_N
\end{pmatrix}
\qquad
\mathbf{V}=\begin{pmatrix}
\mathbf{v}_1^T \\ \vdots\\ \mathbf{v}_N^T
\end{pmatrix}.
\end{align*}

In practice, the frequency shifts are not observable. Instead the frequency difference of arrival (FDOA) is measured between receivers. The FDOA is equivalent to the difference in frequency shifts,
\begin{align}
  \label{eq:fdoa}
  f_{i,j} = d_j-d_i.
\end{align}
A system equivalent to Equation \eqref{eq:fshiftFF} can be constructed for the FDOA, with the use of a differencing matrix $\mbf{P}$. The matrix $\mbf{P}$ has entries of 0 and $\pm 1$ corresponding to the differencing in Equation \eqref{eq:fdoa}. Thus, with the far-field simplification above, the vector of FDOA measurements, $\mbf{f}$, is equivalent to,
\begin{align}
  \label{eq:fdoaFF}
\mbf{f} = -\mbf{PV}\hat{\mbf{x}}.
\end{align}
The matrix $\mbf{-PV}$ will be referred to as $\tilde{\mbf{V}}$ for simplicity.

\begin{figure}[h!]
  \includegraphics[scale=0.7]{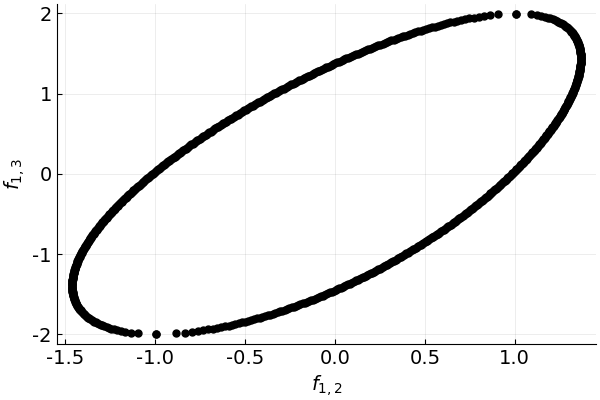}
  \caption{Plot of far-field $f_{1,2}$ vs. $f_{1,3}$ for a system of three receivers centered around the origin. Note the image is an ellipse with scaling in the direction of the left-singular vectors of $\tilde{\mbf{V}}$.}
  \label{f:ellipse}
\end{figure}

This far-field simplification reduces the FDOA equations to a linear system. This suggests that feasible FDOA measurements in the far-field case lie on the image of the unit circle transformed by the matrix $\tilde{\mbf{V}}$. This image is an ellipse with rotation and scaling determined by the singular value decomposition of $\tilde{\mbf{V}}$. Indeed, this can be confirmed by computing the singular value decomposition of generated far-field FDOA measurements and confirming they lie on the same subspace as $\tilde{\mbf{V}}$. This relationship can be demonstrated visually with a plot of generated FDOA measurements (Fig. \ref{f:ellipse}).

\subsection{Calculating direction of arrival (DOA)}
The far-field approximated form of the FDOA equations is linear with variable $\hat{\mbf{x}}$, representing the direction of arrival (DOA) of the signal. Thus, the DOA can be found by solving \eqref{eq:fdoaFF} for $\hat{\mbf{x}}$. If there are more FDOA measurements than direction components, we can find the least squares solution to the problem, which is also the pseudo-inverse solution:
\begin{align}
  \label{eq:doa}
\hat{\mbf{x}} = (\tilde{\mbf{V}}^T\tilde{\mbf{V}})^{-1}\tilde{\mbf{V}}^T\mathbf{f}.
\end{align}

One method for denoising in TDOA-based geolocation is the projection of noisy measurements onto the range of the differencing matrix $\mathbf{P}$~\cite{Schmidt1996,Compagnoni2017}. This ensures that the TDOA measurements are physically realizable and consistent between receivers. One benefit of the method for DOA calculation proposed above is that denoising is automatically performed since projection onto the range of $\mathbf{-PV}$ is equivalent to projection onto the range of $\mathbf{P}$.

%% file: TDOA.tex
\section{Direction of Arrival with TDOA Measurements}
\label{s:TDOA}
\subsection{Far-field approximation for TDOA}
Although the time difference of arrival (TDOA) is simpler than the FDOA case, we include its far-field approximation for completeness.

Using the same problem setup as above, the time it takes for the signal to travel between the emitter and receiver $i$ is:
\begin{align*}
  \tau_{i} = \frac{1}{c}\|\mbf{x_i}-\mbf{x}\|,
\end{align*}
from here the scalar $\frac{1}{c}$ will be left out for simplicity.
The far-field approximation for $\|\mbf{x_i}-\mbf{x}\|$ is given~\cite{Cheney2009},
\begin{align*}
  \|\mbf{x_i}-\mbf{x}\|=\|\mbf{x}\|\left(1-\frac{\mbf{x}_i\cdot\hat{\mbf{x}}}{\|\mbf{x}\|}+\mathcal{O}\left( \frac{\|\mathbf{x}_i\|}{\|\mathbf{x}\|}\right) \right).
\end{align*}
Thus, $\tau_{i}$ becomes,
\begin{align*}
  \tau_{i} = \|\mbf{x}\| - \mbf{x}_i\cdot\hat{\mbf{x}}.
\end{align*}
As in the FDOA case, $\tau_i$ is not observable. Instead we look to the time difference of arrival (TDOA) between receivers $i$ and $j$,
\begin{align*}
  \tau_{i,j} =& \; \left(\|\mbf{x}\| - \mbf{x}_j\cdot\hat{\mbf{x}}\right) - \left(\|\mbf{x}\| - \mbf{x}_i\cdot\hat{\mbf{x}}\right) \\
            =& \; \mbf{x}_i\cdot\hat{\mbf{x}} - \mbf{x}_j\cdot\hat{\mbf{x}} \\
            =& \; (\mbf{x}_i-\mbf{x}_j)\cdot\hat{\mbf{x}}.
\end{align*}
The system of TDOA measurements are equivalent to:
\begin{align}
  \pmb{\tau} = -\mbf{PX}\hat{\mbf{x}},
\end{align}
where $\mbf{X}$ is the matrix of receiver locations and $\mbf{P}$ is a differencing matrix as before. This suggests that feasible far-field TDOA measurements lie in the image of the unit circle under transformation of $-\mbf{PX}$.

\subsection{Calculating direction of arrival (DOA)}
As in the FDOA case, the least-squares estimate of direction of arrival can be calculated using the pseudoinverse:
\begin{align}
    \label{eq:doa_t}
  \hat{\mbf{x}} = -((\mbf{PX})^T\mbf{PX})^{-1}(\mbf{PX})^T\mathbf{\tau}.
\end{align}

%% file: numericalresults.tex
\section{Numerical Results}
\label{s:numerics}

One method of estimator evaluation is the comparison of estimator variance with the Cramer Rao lower bound (CRLB). Assuming data containing noise distributed Gaussian with a given covariance matrix, the CRLB provides a lower bound on the variance of estimator accuracy. We consider here the FDOA-based DOA estimation problem. Consider FDOA measurements, $\hat{f}_{i,j}$, equal to the sum of the true FDOA and Gaussian-distributed deviation. That is,
\begin{align*}
\hat{\mathbf{f}} = \mathbf{f} + \delta \mathbf{f},
\end{align*}
where $E\left[\delta \mathbf{f}\right]=0$ and $E\left[\delta\mathbf{f}\delta\mathbf{f}^T\right]=\mathbf{Q}$. The CRLB can then be computed for data corresponding with covariance matrix $\mathbf{Q}$. This provides a lower bound on variance of DOA estimation using FDOA measurements. It follows that an algorithm with variance near the CRLB has optimal accuracy with the given level of noise. For ease of visualization, we will consider the CRLB corresponding to the AOA (given by $\theta$) as opposed to DOA.

The CRLB of an unbiased estimator is the inverse of the Fisher information matrix, $\mbf{J}$. For the FDOA based AOA problem, this is given by~\cite{Ho1997}:
\begin{align*}
\mbf{J}(\mbf{x},\mbf{X},\mbf{V}; \mbf{Q}) = \left(\frac{\partial\mbf{f}^{T}}{\partial\mbf{x}}\cdot\frac{\partial\mbf{x}}{\partial\theta}\right)\mbf{Q}^{-1}\left(\frac{\partial \mbf{f}}{\partial\mbf{x}^T}\cdot\frac{\partial\mbf{x}^T}{\partial\theta}\right).
\end{align*}
 This can be calculated for a theoretical set of receiver positions ($\mbf{X}$), velocities ($\mbf{V}$), covariance matrix ($\mbf{Q}$), and emitter location ($\mbf{x}$). The result is a single value whose inverse is the CRLB for AOA.

%% Somewhere here we need to include a measure of how 'far-field' we are. One good measure might be the mean of |x_i| divided by |x|. Then we can maybe include a couple of these graphs for one really far field and one near-er field. Then we can include a short discussion of this.
\begin{figure}
  \includegraphics[scale=0.8]{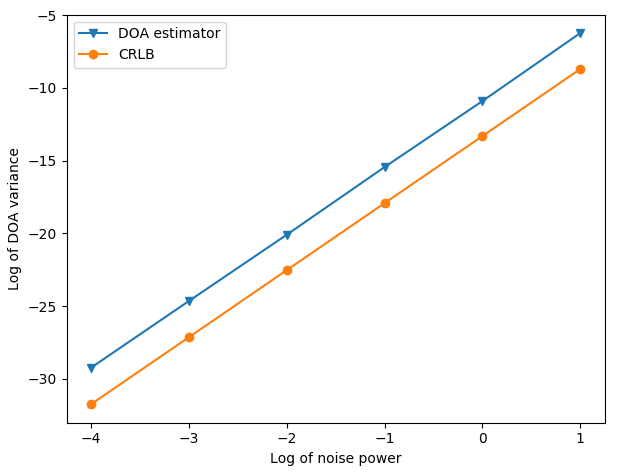}
  \caption{Log-log plot of DOA estimator error vs. the Cramer Rao lower bound on FDOA-based DOA variance. }
  \label{CRLB}
\end{figure}

\begin{figure}
  \includegraphics[scale=0.3]{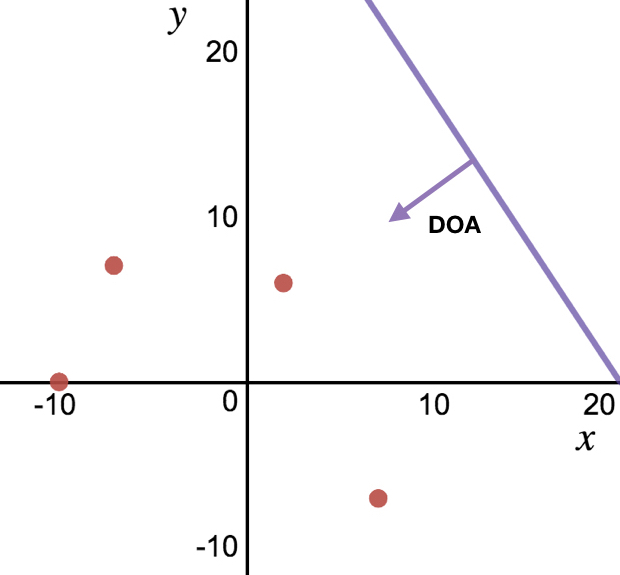}
  \caption{Receiver configuration for CRB comparison in Fig.~\ref{CRLB}.}
  \label{config}
\end{figure}

Numerical trials can then be run with our DOA approximation method and the variance in DOA can be compared to the CRLB. This is the content of Figure \ref{CRLB}. The $x$-axis gives varying levels of noise power and the $y$-axis shows corresponding AOA variance for our approximation and the CRLB. It is clear that the AOA variance trend mimics that of the CRLB.

%% file: conclusion.tex
\section{Conclusion}
\label{s:conclusion}
Considering far-field FDOA-based geolocation naturally leads to a simple method for determining direction of arrival. This calculation requires only a linear solve which makes the corresponding source-localization technique very efficient. Additionally, since FDOA measurement data is projected onto the range of the differencing matrix, the solution is naturally de-noised in a method consistent with~\cite{Schmidt1996,Compagnoni2017}. Another benefit of this method is the generality that allows DOA to be calculated with either TDOA or FDOA measurements. This allows for accurate source localization in the presence of a range of waveforms.